\documentclass[natbib=false,sigconf]{acmart}

\RequirePackage[
  datamodel=acmdatamodel,
  style=acmnumeric,
  ]{biblatex}

\usepackage{makecell}
\usepackage{multirow}
\usepackage[utf8]{inputenc} % allow utf-8 input
\usepackage[T1]{fontenc}    % use 8-bit T1 fonts
\usepackage{threeparttable}
\usepackage{hyperref}       % hyperlinks
\usepackage{url}            % simple URL typesetting
\usepackage{booktabs}       % professional-quality tables
\usepackage{amsfonts}       % blackboard math symbols
\usepackage{nicefrac}       % compact symbols for 1/2, etc.
\usepackage{microtype}      % microtypography
\usepackage{xcolor}         % colors
\usepackage{soul}

\usepackage{graphicx}
\usepackage{bbm}
\usepackage[linesnumbered,ruled,vlined]{algorithm2e}

\usepackage{caption}
\usepackage{textcomp}
\usepackage{amsmath}
\usepackage{enumitem}
\usepackage{wrapfig}
\usepackage{subcaption}

\newcommand{\model}{\textsc{GUME}}

\begin{document}

% \title{Resolving the Long-tail Distribution in Multimedia Recommendation via Multi-view Graph}
% \title{Utilizing Behavior Distillation and Item Similarity to Enhance Multi-media Recommendation System}
% \title{Distilling User Behavior and Leveraging Item Similarity for Enhanced Multi-media Recommendations}
\title{GUME: Graphs and User Modalities Enhancement for Long-Tail Multimodal Recommendation}
% \title{Fusing Behavioral Essence and Item Relatedness: A Dual-Pronged Approach to Elevate Multi-media Recommenders}
% \title{Refining Multi-media Recommendations through Behavior Condensation and Item Kinship}
% \title{Behavior-Item Synergy: An Innovative Fusion for Augmenting Multi-media Recommendation Performance}
% \title{Harnessing the Power of Behavior Compression and Item Similarity for Superior Multi-media Suggestions}
% \title{Multi-media Recommendation Renaissance: Integrating Behavior Quintessence and Item Resemblance}
% \title{Melding Behavior Refinement and Item Congruity for Next-Gen Multi-media Recommendation Engines}
% \title{Distilled Behaviors and Item Affinity: A Recipe for Multi-media Recommendation Excellence}
% \title{Unleashing the Potential of Multi-media Recommenders via Behavior Reduction and Item Parallelism}
% \title{A Symphony of Behavior Essence and Item Kinship for Elevated Multi-media Recommendations}
\author{Guojiao Lin$^{1,2}$, Zhen Meng$^{1,2,*}$, Dongjie Wang$^{3}$, Qingqing Long$^{1}$, Yuanchun Zhou$^{1}$, \\Meng Xiao$^{1,*}$}\thanks{Zhen Meng (zhenm99@cnic.cn) and Meng Xiao (shaow@cnic.cn) are the corresponding authors.}
\affiliation{%
    \institution{$1$ Computer Network Information Center, Chinese Academy of Sciences}
	\institution{$2$ University of the Chinese Academy of Sciences}
    \institution{$3$ University of Kansas}
    \country{}
}

\renewcommand{\shortauthors}{Anonymous Authors}


\begin{abstract}
Multimodal recommendation systems (MMRS) have received considerable attention from the research community due to their ability to jointly utilize information from user behavior and product images and text. Previous research has two main issues. First, many long-tail items in recommendation systems have limited interaction data, making it difficult to learn comprehensive and informative representations. However, past MMRS studies have overlooked this issue. Secondly, users' modality preferences are crucial to their behavior. However, previous research has primarily focused on learning item modality representations, while user modality representations have remained relatively simplistic.
To address these challenges, we propose a novel \underline{G}raphs and \underline{U}ser \underline{M}odalities \underline{E}nhancement ($\model$) for long-tail multimodal recommendation. Specifically, we first enhance the user-item graph using multimodal similarity between items. This improves the connectivity of long-tail items and helps them learn high-quality representations through graph propagation. Then, we construct two types of user modalities: explicit interaction features and extended interest features. 
By using the user modality enhancement strategy to maximize mutual information between these two features, we improve the generalization ability of user modality representations. Additionally, we design an alignment strategy for modality data to remove noise from both internal and external perspectives. 
Extensive experiments on four publicly available datasets demonstrate the effectiveness of our approach. 
The code and data are publicly accessible via ~\href{https://github.com/NanGongNingYi/GUME}{GitHub}.
\end{abstract}
\maketitle

\section{Introduction}
In the era of information explosion, recommendation systems ~\cite{sarwar2001item} have become indispensable tools to help users discover relevant and interesting items from vast amounts of data. 
Among various recommendation methods, those that focus on capturing collaborative signals from user-item interactions have received significant attention. 
However, these methods often suffer from the problem of data sparsity. 
Due to the severe scarcity of interaction data, where 80\% of interactions are focused on popular items, responses for tail items are significantly limited. 
Over time, this not only exacerbates the cold start problem but also traps users in information cocoons. 
Multimodal recommendation systems (MMRS), capable of integrating substantial information across different item modalities, show potential in mitigating this problem and have consequently garnered considerable interest within the research community. 
By leveraging multimodal data, including text and images, MMRS can achieve a deeper insight into the features of the items and user preferences ~\cite{chen2017attentive, deldjoo2016content, wei2017collaborative}, thereby enhancing the precision and variety of their recommendations.

Several studies have integrated multimodal content into recommendation systems. 
For instance, VBPR ~\cite{he2016vbpr} improves item representation by merging visual embeddings with ID embeddings. The application of Graph Convolutional Networks (GCN) to uncover hidden information among users and items has also received increased focus. 
In addition, MMGCN ~\cite{wei2019mmgcn} develops modality-specific bipartite graphs for users and items and combines various modality features for prediction. 
LATTICE ~\cite{zhang2021LATTICE} and MICRO ~\cite{zhang2022micro} generate multi-view semantic graphs based on multi-modal data and then merge these graphs to identify potential item relationship graphs. 
MENTOR ~\cite{xu2024mentor} constructs static item homogeneity graphs for each modality to strengthen semantic relationships between items. Nevertheless, these approaches do not address the use of multimodal information to improve the connectivity of tail items in the user-item interaction graph. 
Due to the sparse interaction data, the tail items receive insufficient information during the graph propagation phase, hindering their ability to develop comprehensive and informative representations. 
Thus, \textbf{it is crucial to take advantage of the multimodal information to improve the connectivity of the graph}, which in turn helps mitigate the cold start issue for long-tail items.

Although item modality information is rich, user modality representation still has much to explore. 
For example, BM3 ~\cite{zhou2023bootstrap} only learns user ID representations and ignores user modality representations. 
SLMRec ~\cite{tao2022SLMRec} and MGCN ~\cite{yu2023MGCN} represent the modality features of the user simply by aggregating the modality features of the items with which the user has interacted. 
MENTOR customizes user modality embeddings, combines them with item modality embeddings, and updates them through GCN propagation. 
However, using simple aggregation or customized methods to represent user modalities cannot effectively capture user modality preferences. 
The simple aggregation method limits user modality preferences to past behavior, while customized methods ignore past preferences and can add noise during propagation. Therefore, \textbf{it is necessary to further explore more effective methods to capture user preference for the user modality}.

To address the above issues, we propose a novel \underline{G}raphs and \underline{U}ser \underline{M}odalities \underline{E}nhancement ($\model$) for long-tail multimodal recommendation. Firstly, we construct modality item graphs based on multimodal similarity and identify the semantic neighbors of items. Then, we add edges between these items and their semantic neighbors to the user-item interaction graph to enhance graph connectivity~\cite{qq1,qq2,qq3}. Next, based on the modality item graphs and the enhanced user-item graph, we extract explicit interaction features~\cite{fs1,fs2,fs3} and extended interest features~\cite{ft1,ft2,ft3,ft4,ft5}, representing the user's historical modality preferences and potential future modality preferences, respectively. 
By leveraging common information between modalities, we separate coarse-grained attributes from explicit interaction features. 
Since fine-grained attributes are related to user behavior, we further use behavior information to reveal fine-grained attributes within the explicit interaction features~\cite{fsa1,fsa2}. 
The separation of these attributes is aimed at better aggregating them, so we reaggregate the coarse-grained and fine-grained attributes to form enhanced explicit-interaction features. 
Then, we design a user-modality enhancement module that maximizes the mutual information between explicit interaction features and extended interest features, improving the generalizability of user modality representations. Additionally, we design an alignment module to capture commonalities within internal information (e.g., image and text) and external information (e.g., behavior and modality), thereby removing noise unrelated to recommendations.

The main contributions of this paper are summarized as follows:
\begin{itemize}
    \item We propose a strategy to enhance user-item graphs based on multimodal similarity, improving the connectivity of tail items.
    \item We develop a user modality enhancement strategy that improves the generalization ability of user modality representations, enabling them to effectively adapt to new products or changes in user behavior, even without direct interaction data.
    \item We design an alignment strategy from internal and external perspectives to capture commonalities within modalities as well as between modalities and external behaviors, thereby achieving a denoising effect.
    \item We conduct comprehensive experiments on four public Amazon datasets to demonstrate the unique advantages of our $\model$.
\end{itemize}

\section{Problem Definition}
Let \(\mathcal{U} = \{u\}\) denote the user set and \(\mathcal{I}=\{i\}\) denote the item set. The ID embeddings for users and items are represented as \(E_{id} = \{E_{u,id} \| E_{i,id}\} \in \mathbb{R}^{d \times (|\mathcal{U}| + |\mathcal{I}|)}\), where \(d\) is the embedding dimension. The modality embeddings for users are \(E_{u,m} \in \mathbb{R}^{d \times |\mathcal{U}|}\), and the modality features for items are \(E_{i,m} \in \mathbb{R}^{d_m \times |\mathcal{I}|}\). Here, \(m \in \mathcal{M}\) is the modality, \(\mathcal{M} = \{v, t\}\) is the set of modalities, and \(d_m\) is the dimension of the features. In this paper, we primarily consider visual and textual modalities, but our method can also be extended to additional modalities.

The matrix \(\mathcal{R} \in \{0,1\}^{|U|\times|I|}\) represents user-item interactions, where \(\mathcal{R}_{u,i}=1\) indicates that user \(u\) clicked on item \(i\), and \(\mathcal{R}_{u,i}=0\) otherwise. Based on the interaction matrix \(\mathcal{R}\), we construct a bipartite graph \(\mathcal{G}=(\mathcal{V}, \mathcal{E})\), where \(\mathcal{V}=\mathcal{U}\cup\mathcal{I}\) represents the set of nodes, and \(\mathcal{E}=\{(u, i) \mid u \in \mathcal{U}, i \in \mathcal{I}, \mathcal{R}_{u,i}=1\}\) represents the set of edges. The primary goal of multimodal recommendation systems is to recommend the top-N most relevant items to each user based on the predicted preference score \(y_{ui} = f_{\Theta}(e_{id}, e_{u,m}, e_{i,m})\), where \(\Theta\) represents the model parameters.

\section{Methodology}
\begin{figure*}
  \centering
  \includegraphics[width=\textwidth]{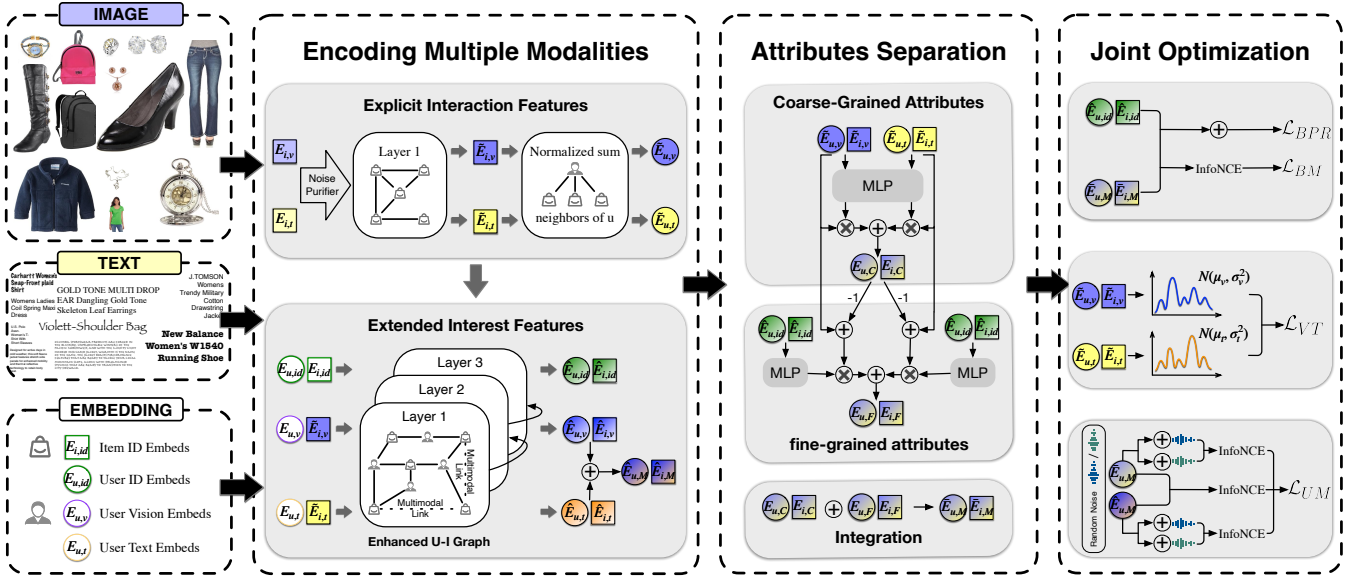}
  \caption{The overview of $\model$. We first utilize a graph convolutional network to extract explicit interaction features and extended interest features. Then, we separate and aggregate the attributes of the explicit interaction features to achieve denoising. We maximize the mutual information between explicit interaction features and extended interest features. Finally, we align information within internal modalities as well as between modalities and external behaviors.}
  \label{fig:model}
\end{figure*}

\subsection{Enhancing User-Item Graph}
Inspired by the successful application of graph enhancement learning in long-tail recommendation tasks \cite{luo2023improving}, we have introduced this approach to multimodal recommendation. Specifically, we first construct modality item graphs based on multimodal similarities between items. Subsequently, we select items that are similar to the target item in both textual and visual attributes as its semantic neighbors. Finally, we add the edges between the target items and their semantic neighbors to the user-item graph. This strategy helps improve the connectivity of tail items.
\subsubsection{\textbf{Constructing Modality Item Graphs}}
We employ the KNN algorithm \cite{chen2009knn} to construct an item-item graph for each modality m, aimed at extracting multimodal relationships between items. Specifically, we calculate the similarity score $ S^m_{i,j} $ for the item pair $(i,j)\in I$ by measuring the cosine similarity between their original modality features, $ e^m_i $ and $ e^m_j $.
\begin{equation}
    S^m_{i,j} = \frac{(e^m_i)^Te^m_j}{\left \| e^m_i \right \| \left \| e^m_j \right \|} 
\end{equation}
We retain only the top-k neighbors with the highest similarity scores to capture the most relevant features:
\begin{equation}
    \bar{S}^m_{i,j}=\left\{\begin{matrix}
                     S^m_{i,j}, & S^m_{i,j}\in top-K(\{S^m_{i,p},p\in I\}),\\ 
                     0, & otherwise, 
                    \end{matrix}\right.
\end{equation}
where $\bar{S}^m_{i,j}$ represents the edge weight between item \(i\) and item \(j\) within modality \(m\), and $\{S^m_{i,p},p\in I\}$ represents the neighbor scores for the i-th item. To mitigate the issues of gradient explosion or vanishing, we normalize the adjacency matrix as follows:
\begin{equation}
    \tilde{S}^m = (D^m)^{-\frac{1}{2}}\bar{S}^m(D^m)^{-\frac{1}{2}}
\end{equation}
where $D^m$ is the diagonal degree matrix of $\bar{S}^m$. Inspired by \cite{zhou2023tale}, we freeze each modality item graph after initialization.

\subsubsection{\textbf{Identifying Semantic Neighbors}}
According to \cite{liu2021interest}, in recommendation systems based on Graph Convolutional Networks (GCN), the indiscriminate use of high-order nodes can introduce negative information during the embedding propagation process. This is particularly problematic when the model stacks more layers, as it can lead to performance degradation. Therefore, indiscriminate graph augmentation is inadvisable, as it may lead to the transmission of irrelevant information between items. It is important to identify semantic neighbors that are not only similar in features to the target item but are also likely to have meaningful connections.

To address this issue, we introduce a strategy based on multimodal similarity to identify semantic neighbors. Specifically, this is implemented by utilizing the modality item graph. This graph keeps only the top-k neighbors with the highest similarity scores for each item, and we use it to identify items that are similar to the target item across multiple modalities (textual and visual). We then define these items as the semantic neighbors of the target item. The set of items and their semantic neighbors can be expressed as:
\begin{equation}
    \mathcal{C} = \cup_{i \in \mathcal{I}} \{(i, j) \mid j \in \mathbb{N}_i\}
\end{equation}
where, \(\mathbb{N}_i\) represents the semantic neighbors of item \(i\). Then, we enhance the user-item graph by adding edges between items and their semantic neighbors. Formally,
\begin{equation}
    \mathcal{A} = \begin{pmatrix}
                      0   &  R \\ 
                      R^T &  \mathcal{C}
                  \end{pmatrix}
\end{equation}
where 0 is the all-zero matrix, and R is the user-item interaction matrix.

\subsection{Encoding Multiple Modalities}
According to \cite{xu2024mentor, yu2023MGCN}, user-item graphs and modality item graphs contain rich collaborative and semantic signals that can significantly enhance the performance of multimodal recommendation systems. Inspired by them, we leverage modality item graphs to extract explicit interaction features, while also using enhanced user-item graphs to extract extended interest features. These features represent the user's historical modality preferences and potential future modality preferences, respectively.

\subsubsection{\textbf{Extracting Explicit Interaction Features}}
We first design a feature space transformation function that maps the initial modality features to the same space as the ID features:
\begin{equation}
    \dot{f}_m(X) = \sigma(\mathcal{W}_2(\mathcal{W}_1X+b_1)+b_2)
\end{equation}
where $\mathcal{W}_1 \in \mathbb{R}^{d\times d_m}$ and $\mathcal{W}_2 \in \mathbb{R}^{d\times d}$ denote the linear transformation matrixs, $b_1 \in \mathbb{R}^d$ and $b_2 \in \mathbb{R}^d$ denote the bias vectors and $\sigma$ is the sigmoid function. These parameters are modality-specific.

Then, inspired by \cite{yu2023MGCN}, we perform an element-wise product between the transformed modality features and the ID features to remove noise unrelated to behavior from the modality features.
\begin{equation}
    \dot{E}_{i,m} = E_{i,id}\odot \dot{f}_m(E_{i,m})
\end{equation}
where $\odot$ represents the element-wise product.

After obtaining the denoised modality features, we perform graph convolution operations on the modality item graphs to propagate and update the item modality features.
\begin{equation}
    \tilde{E}_{i,m} = \tilde{S}^m\dot{E}_{i,m}
\end{equation}

Next, we aggregate the modality features of the user's neighboring items to represent the user's explicit modality features.
\begin{equation}
    \tilde{E}_{u,m} = \mathcal{R}\tilde{E}_{i,m}
\end{equation}
Where $\mathcal{R}\in \{0,1\}^{|U|\times|I|}$ represents the user-item interaction matrix.

The ultimate explicit interaction features are derived by combining the user's explicit modality feature \(\tilde{E}_{u,m}\) with the item modality features \(\tilde{E}_{i,m}\). Formally,
\begin{equation}
    \tilde{E}_{m} = \{\tilde{E}_{u,m}||\tilde{E}_{i,m}\}
\end{equation}
where $||$ denotes the concatenation operation.

\subsubsection{\textbf{Extracting Extended Interest Features}}
Let \(\hat{E}_{u,m}\) and \(\hat{E}_{i,m}\) represent the stacked extended interest embeddings for users and items, respectively. We propagate the specific modality embeddings on the enhanced user-item graph to extract modality-specific extended interest embeddings. The same applies to ID embeddings. Specifically, at the \(l^{th}\) layer of graph convolution, the embeddings of users and items can be represented as:
\begin{equation}
\begin{split}
    \hat{E}^{(l)}_m = (D^{-\frac{1}{2}} \mathcal{A} D^{-\frac{1}{2}})\hat{E}^{(l-1)}_m
    % \\[6pt]
    % \hat{E}^{(l)}_{id} = (D^{-\frac{1}{2}} \mathcal{A} D^{-\frac{1}{2}})\hat{E}^{(l-1)}_{id}
\end{split}
\end{equation}
Where \( \hat{E}^{(l)}_m \) represent the specific modality embeddings for users and items at the \( l \)-th layer of graph convolution, \( D \) is the diagonal degree matrix of \( \mathcal{A} \). \( \hat{E}^{(0)}_m \) is the concatenation of \(E_{u,m}\) and \( \tilde{E}_{i,m} \).

By aggregating multi-layer neighbor information, we obtain the final modality-specific extended interest features.
\begin{equation}
\begin{split}
    % \hat{E}_{id} = \frac{1}{L+1}\sum_{i=0}^{L}\hat{E}^{(l)}_{id}\\
    \hat{E}_{m} = \frac{1}{L+1}\sum_{i=0}^{L}\hat{E}^{(l)}_{m}
\end{split}
\end{equation}
where L is the number of user-item graph layers. Finally, we fuse the extended interest features from the visual and textual modalities to obtain the final extended interest features:
\begin{equation}
    \hat{E}_M = \sum_{m\in \mathcal{M}}\hat{E}_{m}
\end{equation}

\subsection{Attributes Separation for Better Integration}
Multiple modalities convey comprehensive information \cite{baltruvsaitis2018multimodal}. For example, images and text of clothing items can both reflect the coarse-grained attribute~\cite{m1,m2,m3} of the category. However, images can provide fine-grained attributes specific to visuals, such as patterns and styles, while text can provide fine-grained attributes specific to text, such as fabric and size. The coarse-grained attributes represent commonalities among items, whereas fine-grained attributes are key factors influencing users purchasing decisions. 
Since the explicit interaction features are only enhanced through the modality item graph and represent user modality by simple aggregation, they lack information related to user behavior. Therefore, we first separate the coarse-grained attributes in the explicit interaction features, and then use user behavior information to reveal the fine-grained attributes within it.

\subsubsection{\textbf{Separating Coarse-Grained Attributes}} We design a multilayer perceptron (MLP) to process the input vector X, apply a nonlinear transformation to X, and then output it as a weight matrix:
\begin{equation}
    f(X) = \mathcal{W}_4tanh(\mathcal{W}_3X+b_3)    
\end{equation}
where $\mathcal{W}_4\in \mathbb{R}^d$ denotes attention vector and $\mathcal{W}_3\in \mathbb{R}^{d\times d}$, $b_3\in \mathbb{R}^d$ denote the weight matrix and bias vector, respectively. These parameters are shared for all modalities.

We input $\tilde{E}_m$ into MLP to get the output, which is then passed through softmax to obtain the importance scores for each modality. Then, the coarse-grained attributes can be represented as:
\begin{equation}
    E_{C} = \sum_{m\in \mathcal{M}}\frac{exp(f(\tilde{E}_m))}{\sum_{{m}'\in \mathcal{M}}exp(f(\tilde{E}_{m}'))}\tilde{E}_m
\end{equation}

\subsubsection{\textbf{Separating Fine-Grained Attributes}} We designed a modality-specific behavior information extraction function \(\ddot{f}_m(X)\) as follows:
\begin{equation}
    \ddot{f}_m(X) = \sigma(\mathcal{W}_5X+b_5)
\end{equation}
where $\mathcal{W}_5\in \mathbb{R}^{d\times d}$ and $b_5\in \mathbb{R}^d$ denote the weight matrix and bias vector. These parameters are modality-specific.

To extract fine-grained attributes, we first subtract the coarse-grained attributes \(E_{C}\) from \(\tilde{E}_m\), then multiply it by \(\hat{E}_{id}\) processed by the function \(\ddot{f}_m(X)\). Finally, we sum the fine-grained attributes from all modalities to obtain the final representation.
\begin{equation}
    E_{F} = \frac{1}{|M|}\sum_{m\in \mathcal{M}}(\tilde{E}_m - E_{C}) \odot\ddot{f}_m(\hat{E}_{id})
\end{equation}

Ultimately, we integrate coarse-grained attributes $E_{C}$ with fine-grained attributes $E_{F}$ as the final enhanced explicit interaction features $\bar{E}_{M}$:
\begin{equation}
    \bar{E}_M = E_{C} + E_{F}
\end{equation}

\subsection{Capturing Commonalities Through Alignment}
In recommendation scenarios, users may be attracted to common features between the images and text of products, while there are also potential correlations between behavioral features and modal features. To capture the commonalities between different modalities, we designed two alignment tasks from different perspectives.

Specifically, we first capture common information from an internal perspective. Inspired by \cite{xu2024mentor}, we parameterize the visual modality \(\tilde{E}_v\) and the textual modality \(\tilde{E}_t\) using Gaussian distributions.
\begin{equation}
     \tilde{E}_v\sim N(\mu_v,\sigma^2_v), 
     \quad
     \tilde{E}_t\sim N(\mu_t,\sigma^2_t), 
\end{equation}
where $(\mu_v,\sigma^2_v)$ and $(\mu_t,\sigma^2_t)$ represent the Gaussian distribution parameters for \(\tilde{E}_v\) and \(\tilde{E}_t\). By minimizing the differences between the means and standard deviations of these two modalities, we enhance the internal connections between the modalities. The formula is as follows:
\begin{equation}
     \mathcal{L}_{VT} = |\mu_v - \mu_t| + |\sigma_v - \sigma_t|
\end{equation}

Then, we capture common information from an external perspective. We encourage the explicit interaction features of users and items that have similar interaction behaviors to be close to each other, using InfoNCE to align \(\hat{E}_{id}\) and \(\bar{E}_M\):

\begin{equation}
\begin{aligned}
     \mathcal{L}_{BM} = & \sum_{u\in \mathcal{U}}-log\frac{exp(\hat{e}_{u,id}\cdot \bar{e}_{u,M}/\mathcal{\tau})}{\sum_{v\in \mathcal{U}}exp(\hat{e}_{v,id}\cdot \bar{e}_{v,M}/\mathcal{\tau})} \\
     & + \sum_{i\in \mathcal{I}}-log\frac{exp(\hat{e}_{i,id}\cdot \bar{e}_{i,M}/\mathcal{\tau})}{\sum_{j\in \mathcal{I}}exp(\hat{e}_{j,id}\cdot \bar{e}_{j,M}/\mathcal{\tau})}
\end{aligned}
\end{equation}
where $\mathcal{\tau}$ is the temperature hyper-parameter of softmax.

Finally, the total alignment loss for capturing commonalities is calculated as:
\begin{equation}
    \mathcal{L}_{AL} = \alpha\mathcal{L}_{VT} + \beta\mathcal{L}_{BM}
\end{equation}
where $\alpha$ and $\beta$ are the balancing hyper-parameters.

\subsection{Enhancing User Modality Representation}
\(\bar{E}_{u,M}\) directly reflects the user's historical interactions, clearly expressing the user's past preferences. However, this is also its limitation, as it lacks foresight into the user's potential interests. In contrast, \(\hat{E}_{u,M}\) not only includes the user's historical modality information but also expands their interests by considering items similar to their historical items. This extension takes into account that users might be interested in items similar to those they have interacted with before, helping to foresee their potential interests.

Based on this idea, we use InfoNCE to maximize the mutual information between \(\bar{E}_{u,M}\) and \(\hat{E}_{u,M}\), enabling \(\bar{E}_{u,M}\) to absorb and integrate the information provided by \(\hat{E}_{u,M}\). This strategy enhances the generalization ability of \(\bar{E}_{u,M}\), enabling it to effectively adapt to new products or changes in user behavior even in the absence of direct interaction data. The mathematical expression of this task is as follows:
\begin{equation}
     \mathcal{L}_C = \sum_{u\in \mathcal{U}}-log\frac{exp(\bar{e}_{u,M}\cdot \hat{e}_{u,M}/\mathcal{\tau})}{\sum_{v\in \mathcal{U}}exp(\bar{e}_{v,M}\cdot \hat{e}_{v,M}/\mathcal{\tau})}
\end{equation}
where $\mathcal{\tau}$ is the temperature hyper-parameter of softmax.

Then, to avoid over-relying on specific features during the learning process and learn user interests from a broader feature distribution, we follow the approach of SimGCL \cite{yu2022SimGCL}. We add uniform noise to \(\bar{E}_{u,M}\) and \(\hat{E}_{u,M}\) to create contrastive views. The process of adding noise is as follows:
\begin{equation}
\begin{split}
     {\bar{e}_{u,M}}' = \bar{e}_{u,M} + {\Delta}',
     \quad
     {\bar{e}_{u,M}}'' = \bar{e}_{u,M} + {\Delta}''
     \\[6pt]
     {\hat{e}_{u,M}}' = \hat{e}_{u,M} + {\Delta}',
     \quad
     {\hat{e}_{u,M}}'' = \hat{e}_{u,M} + {\Delta}''    
\end{split}
\end{equation}
where \({\Delta}'\) and \({\Delta}''\) are random noise vectors drawn from a uniform distribution \(U(0,1)\). To optimize the noise-processed vectors, we define two new loss functions, \(\mathcal{L}_{\bar{N}}\) and \(\mathcal{L}_{\hat{N}}\), which are optimized using the InfoNCE loss function:
\begin{equation}
\begin{split}
    \mathcal{L}_{\bar{N}} = \sum_{u\in \mathcal{U}}-log\frac{exp({\bar{e}_{u,M}}'\cdot{\bar{e}_{u,M}}''/\mathcal{\tau})}{\sum_{v\in \mathcal{U}}exp({\bar{e}_{v,M}}'\cdot {\bar{e}_{v,M}}''/\mathcal{\tau})}
     \\[6pt]
    \mathcal{L}_{\hat{N}} = \sum_{u\in \mathcal{U}}-log\frac{exp({\hat{e}_{u,M}}'\cdot{\hat{e}_{u,M}}''/\mathcal{\tau})}{\sum_{v\in \mathcal{U}}exp({\hat{e}_{v,M}}'\cdot {\hat{e}_{v,M}}''/\mathcal{\tau})}
\end{split}
\end{equation}
where $\mathcal{\tau}$ is the temperature hyper-parameter of softmax.

Finally, the total user modality enhancement loss is calculated as:
\begin{equation}
    \mathcal{L}_{UM} = \gamma(\mathcal{L}_C + \mathcal{L}_{\bar{N}} + \mathcal{L}_{\hat{N}})
\end{equation}
where $\gamma$ is the balancing hyper-parameters.

\subsection{Model Prediction}
Based on the ID features and enhanced explicit interaction features, we can get the general representations of users and items:
\begin{equation}
    e_u = \hat{e}_{u,id} + \bar{e}_{u,M}
\end{equation}
\begin{equation}
    e_i = \hat{e}_{i,id} + \bar{e}_{i,M}
\end{equation}
Finally, we compute the inner product of user and item representations to predict their compatibility score:
\begin{equation}
     y_{u,i}=e_u^Te_i
\end{equation}
\subsection{Optimization}
To optimize the effectiveness of our recommendations, we employ the Bayesian Personalized Ranking (BPR) loss \cite{rendle2012bpr} as our primary optimization function. This approach assumes that the predicted scores for observed user-item pairs should be higher than those for unobserved pairs. The BPR function is defined as follow:
\begin{equation}
     L_{bpr}=\sum_{u,i,j\in \mathcal{D}}-log\sigma(y_{u,i}-y_{u,j})
\end{equation}
where \( D \) represents the training set, \( (u, i) \) are observed user-item pairs and \( (u, j) \) are unobserved pairs. \(\sigma\) is the sigmoid function.

Ultimately, we update the representations of users and items by combining the following loss functions: bpr loss, alignment loss and user modality enhancement loss. The formula is as follows:
\begin{equation}
     \mathcal{L}_{\model} = L_{bpr} + \mathcal{L}_{AL} + \mathcal{L}_{UM} + \delta \left\|\Theta\right\|^2_2
\end{equation}
where $\delta$ is a hyperparameter to control the effect of the $L_2$ regularization, and \( \Theta \) is the set of model parameters.

\begin{table}[htbp]
    \centering
    \caption{Statistics of the experimental datasets}
    \label{ds_stat}
    \resizebox{\linewidth}{!}{
    \begin{tabular}{ccccc}
    \toprule
    Dataset & \#User & \#Item & \#Behavior & Unpopularity \\
    \midrule
         Baby        & 19,445  & 7,050  & 160,792   & 97.83\%\\
         Sports      & 35,598  & 18,357 & 296,337   & 97.62\%\\
         Clothing    & 39,387  & 23,033 & 278,677   & 96.32\%\\
         Electronics & 192,403 & 63,001 & 1,689,188 & 99.06\%\\
    \bottomrule
    \end{tabular}}
\end{table}

\section{Experiments}
In this section, we construct experiments to demonstrate the advantages of the proposed methods and to address the following research questions:
\textbf{RQ1:} How effective is our $\model$ compared to the state-of-the-art multimodal recommendation methods and traditional recommendation methods?
\textbf{RQ2:} How do the key modules impact the performance of our \model?
\textbf{RQ3:} Does the proposed graph augmentation strategy improve the recommendation performance for tail items?
\textbf{RQ4:} Why is the user modality enhancement modules effective?
\textbf{RQ5:} How do different hyper-parameter settings impact the performance of our \model?

\subsection{Experimental Setting}
\subsubsection{\textbf{Dataset}}
We adhere to the dataset selection criteria utilized in several recent studies ~\cite{yu2023MGCN, zhou2023bootstrap, yu2023ld4mrec}, ensuring consistency and comparability of results. Specifically, we conduct extensive experiments on four categories of the Amazon dataset \footnote{The datasets can be found at \url{http://jmcauley.ucsd.edu/data/amazon/links.html}}: (1) Baby, (2) Sports and Outdoors (referred to as Sports), (3) Clothing, Shoes, and Jewelry (referred to as Clothing), and (4) Electronics. These categories offer a diverse range of items, from basic baby care products to high-tech electronic devices. For each modality feature extraction, we follow the same setting mentioned in ~\cite{zhou2023mmrec}, which extracted 4096 dimensions of visual features and 384 dimensions of textual features via pretrained encoder. The item interaction counts in all datasets exhibit a long-tail distribution, which also makes these datasets suitable for research aimed at improving the recommendation quality for long-tail items. The statistics of these datasets can be found in Table~\ref{ds_stat}.

\begin{table*}[htbp]
\centering
\caption{Performance comparison of Baselines and \model\ in terms of Recall@K (R@K), and NDCG@K (N@K). The best results are highlighted in \textbf{bold}. The second-best results are highlighted in underline. (Higher values indicate better performance.)}
\label{tab:comparison_methods}
\begin{tabular}{@{}c|c|cc|cccccccc@{}}
\toprule
Datasets & Metrics & MF-BPR & LightGCN & VBPR & MMGCN & SLMRec & LATTICE{\color{blue}$^*$} & BM3 & MGCN & MENTOR & \model\ \\
\midrule
\multirow{4}{*}{Baby} 
& R@10 & 0.0357 & 0.0479 & 0.0423 & 0.0378 & 0.0540 & 0.0547 & 0.0564 & 0.0620 & \textbf{0.0678} & \underline{0.0673} \\
& R@20 & 0.0575 & 0.0754 & 0.0663 & 0.0615 & 0.0810 & 0.0850 & 0.0883 & 0.0964 & \textbf{0.1048} & \underline{0.1042} \\
& N@10 & 0.0192 & 0.0257 & 0.0223 & 0.0200 & 0.0285 & 0.0292 & 0.0301 & 0.0339 & \underline{0.0362} & \textbf{0.0365} \\
& N@20 & 0.0249 & 0.0328 & 0.0284 & 0.0261 & 0.0357 & 0.0370 & 0.0383 & 0.0427 & \underline{0.0450} & \textbf{0.0460} \\
\midrule 
\multirow{4}{*}{Sports}
& R@10 & 0.0432 & 0.0569 & 0.0558 & 0.0370 & 0.0676 & 0.0620 & 0.0656 & 0.0729 & \underline{0.0763} & \textbf{0.0778} \\
& R@20 & 0.0653 & 0.0864 & 0.0856 & 0.0605 & 0.1017 & 0.0953 & 0.0980 & 0.1106 & \underline{0.1139} & \textbf{0.1165} \\
& N@10 & 0.0241 & 0.0313 & 0.0307 & 0.0193 & 0.0374 & 0.0335 & 0.0355 & 0.0397 & \underline{0.0409} & \textbf{0.0427} \\
& N@20 & 0.0298 & 0.0387 & 0.0384 & 0.0254 & 0.0462 & 0.0421 & 0.0438 & 0.0496 & \underline{0.0511} & \textbf{0.0527} \\
\midrule
\multirow{4}{*}{Clothing}
& R@10 & 0.0187 & 0.0340 & 0.0423 & 0.0378 & 0.0540 & 0.0492 & 0.0422 & 0.0641 & \underline{0.0668} & \textbf{0.0703} \\
& R@20 & 0.0279 & 0.0526 & 0.0663 & 0.0615 & 0.0810 & 0.0733 & 0.0621 & 0.0945 & \underline{0.0989} & \textbf{0.1024} \\
& N@10 & 0.0103 & 0.0188 & 0.0223 & 0.0200 & 0.0285 & 0.0268 & 0.0231 & 0.0347 & \underline{0.0360} & \textbf{0.0384} \\
& N@20 & 0.0126 & 0.0236 & 0.0284 & 0.0261 & 0.0357 & 0.0330 & 0.0281 & 0.0428 & \underline{0.0441} & \textbf{0.0466} \\
\midrule
\multirow{4}{*}{Electronics}
& R@10 & 0.0235 & 0.0363 & 0.0293 & 0.0207 & 0.0422 & - & 0.0437 & \underline{0.0442} & 0.0439 & \textbf{0.0458} \\
& R@20 & 0.0367 & 0.0540 & 0.0453 & 0.0331 & 0.0630 & - & 0.0648 & 0.0650 & \underline{0.0655} & \textbf{0.0680} \\
& N@10 & 0.0127 & 0.0204 & 0.0159 & 0.0109 & 0.0237 & - & \underline{0.0247} & 0.0246 & 0.0244 & \textbf{0.0253} \\
& N@20 & 0.0161 & 0.0250 & 0.0202 & 0.0141 & 0.0291 & - & \underline{0.0302} & \underline{0.0302} & 0.0300 & \textbf{0.0310} \\
\bottomrule
\end{tabular}
\begin{tablenotes}
    \small
    \item {\color{blue}*} indicates the model cannot be fitted into a NVIDIA V100 (32GB) /A100 (40GB) /A100 (80GB) GPU card. 
    \end{tablenotes}
\label{overall performance}
\end{table*}

\subsubsection{\textbf{Compared Methods}}
We compare \model\ with 9 baselines, including 2 traditional recommendation models and 7 Multimodal recommendation models. The details of those compared methods are listed as follow:

\begin{enumerate}[label=(\alph*), leftmargin=2em, itemsep=-1em]
    \item{\textbf{Traditional Recommendation Models: }}\\
    \textbf{MF-BPR}\cite{rendle2012bpr}: This method optimizes recommendation systems based on matrix factorization techniques by incorporating Bayesian Personalized Ranking (BPR) loss.\\
    \textbf{LightGCN}\cite{he2020lightgcn}: This method simplifies the design of GCN, retaining only the neighborhood aggregation suitable for collaborative filtering.\\
    
    \item{\textbf{Multimodal Recommendation Models: }}\\
    \textbf{VBPR}\cite{he2016vbpr}: This method extends the MF-BPR method by integrating the visual features and ID embeddings of each item as its representation, and inputs these into the matrix factorization framework.\\
    \textbf{MMGCN}\cite{wei2019mmgcn}: This model constructs specific graphs for different modalities and learns the features of users and items through these graphs. It then concatenates all modal features for prediction.\\
    \textbf{SLMRec}\cite{tao2022SLMRec}: This model incorporates SSL (Self-Supervised Learning) into graph neural network-based recommendation models and proposes three data augmentation methods. It aims to uncover latent patterns within the multimodal information.\\
    \textbf{LATTICE}\cite{zhang2021LATTICE}: This model learns the item-item structure for each modality and aggregates them to form a semantic item-item graph, in order to obtain better item representations.\\
    \textbf{BM3}\cite{zhou2023bootstrap}: This model simplifies the self-supervised multimodal recommendation model by adopting a latent representation dropout mechanism in place of graph augmentation for generating contrastive views.\\
    \textbf{MGCN}\cite{yu2023MGCN}: This model purifies modal features with the help of item behavior information, reducing modal noise contamination, and models modal preferences based on user behavior.\\
    \textbf{MENTOR}\cite{xu2024mentor}: This model employs aligned self-supervised tasks to synchronize multiple modalities while preserving interaction information. It enhances features through feature masking tasks and graph perturbation tasks.
\end{enumerate}

\subsubsection{\textbf{Evaluation Metrics}}
To ensure a fair evaluation of performance, we utilize two widely adopted metrics: Recall@K (R@K) and NDCG@K (N@K). Following the popular evaluation setup \cite{yu2023MGCN}, we employ a random data split of 8:1:1 for training, validation, and testing phases. We assess the recommendation performance of various methods by reporting the average metrics for all users in the test set under top-K conditions, with K empirically set at 10 and 20.

\subsubsection{\textbf{Implementation Details}}
We implemented our proposed $\model$ and all baseline models using MMRec \cite{zhou2023mmrec}. To ensure fair comparisons, we fixed the embedding sizes for users and items at 64, initialized embedding parameters using the Xavier method \cite{glorot2010understanding}, and used Adam \cite{kingma2014adam} as the optimizer with a learning rate of $1e^{-3}$. For our proposed $\model$, we performed a grid search on hyper-parameters $\alpha$, $\beta$, and $\gamma$ in \(\{1e^{-4}, 1e^{-3}, 1e^{-2}, 1e^{-1}, 1e^0\}\), and temperature hyper-parameters \(\tau\) in \(\{0.1, 0.2, 0.4, 0.6, 0.8, 1.0\}\). The GCN layer in the User-Item graph was fixed at 3, while the layer in the Item-Item graph was fixed at 1 (except baby is 2). The k for top-k in the Item-Item graph was set at 10. Early stopping and total epochs were fixed at 20 and 1000, respectively. Following \cite{yu2023MGCN}, we used Recall@20 on the validation data as the training-stopping criterion.

\subsection{Overall Performance (RQ1)}
Table \ref{overall performance} shows the performance comparison of the proposed $\model$ and other baseline methods on four datasets. The table reveals the following observations:

Our $\model$ model achieved excellent performance across multiple metrics, surpassing traditional recommendation models and multimodal recommendation models. Specifically, in terms of Recall@20 for Sports, Clothing, and Electronics, $\model$ outperforms the best baseline by $2.28\%$, $3.54\%$, and $3.82\%$ respectively; while in terms of NDCG@20, it shows improvements of $3.13\%$, $5.67\%$, and $2.65\%$. On the Baby dataset, $\model$ ties with the best baseline in Recall@20 and improves by $2.22\%$ over the best baseline in NDCG@20. The results validate the effectiveness of our $\model$.

Using multimodal information of items to enhance graph connectivity can improve recommendation performance. For example, LATTICE dynamically learns the latent structure between items based on the similarity of their multimodal features. However, dynamically generating latent structure is unnecessary and memory-intensive, which is not efficient for computation. This was confirmed by the subsequent work FREEDOM \cite{zhou2023tale}. Inspired by FREEDOM, MENTOR constructs item homogeneous graphs for each modality to enhance semantic relationships between items and then freezes them. Our $\model$ first identifies semantic neighbors of items based on multimodal similarity and then enhances the user-item interaction graph with them, especially by adding more edges for tail items. This promotes the exploration of items that users might be interested in and alleviates the cold start problem for tail items.

Aligning multiple types of features can improve recommendation performance. For example, BM3 aligns modality features from both intra-modality and inter-modality perspectives. MGCN designs a self-supervised auxiliary task to promote the exploration between behaviors and multimodal information. MENTOR designed a multilevel cross-modal alignment task, aligning each modality under the guidance of ID embeddings while maintaining historical interaction information. Our $\model$ not only aligns external information such as behavioral features and multimodal features but also aligns internal information such as image features and text features, promoting the learning of user and item representations from multiple perspectives.

The multimodal representation of users has always been relatively vague. Some methods, such as BM3, only learn user ID representations and ignore users' multimodal representations. This is a neglect of users' multimodal preferences. Other methods represent users' multimodal features by aggregating the multimodal features of items. For example, SLMRec and MGCN obtain users' multimodal features by simply aggregating the multimodal features of interacted items. This approach treats multimodal features of purchased items equally, which is not effective for fully capturing users' multimodal preferences. Building on this, our $\model$ constructs extended interest features for users. By using contrastive learning with users' explicit interaction features, we maximize the mutual information between them. This process helps to refine and expand users' multimodal preferences effectively.

\subsection{Ablation Study (RQ2)}
In our work, \(\model\) comprises the modules Graph Enhancement, Alignment for Capturing Commonalities and User Modality Augment. To thoroughly examine the impact of these modules, we conduct comprehensive ablation studies. We use "w/o XX" to denote the absence of a specific module, meaning "without XX".

\begin{table}[H]
\centering
\caption{Performance comparison between different variants.}
\begin{tabular}{lc|ccccc}
\hline
\multicolumn{1}{c}{Variants} & \multicolumn{1}{c|}{Metrics} & \multicolumn{4}{c}{Datasets} \\ 
\multicolumn{1}{c}{} & & Baby & Sports & Clothing & Electronics \\ 
\hline
\model        & R@20 & 0.1042 & 0.1165 & 0.1024 & 0.0680 \\
              & N@20 & 0.0460 & 0.0527 & 0.0466 & 0.0310 \\
w/o GE        & R@20 & 0.1039 & 0.1155 & 0.1003 & 0.0679 \\
              & N@20 & 0.0460 & 0.0525 & 0.0456 & 0.0310 \\
w/o AL        & R@20 & 0.0954 & 0.1076 & 0.0862 & 0.0642 \\
              & N@20 & 0.0417 & 0.0477 & 0.0387 & 0.0290 \\
w/o UM        & R@20 & 0.1011 & 0.1127 & 0.0988 & 0.0656 \\
              & N@20 & 0.0449 & 0.0509 & 0.0450 & 0.0303 \\
\hline
\end{tabular}
\label{ablation_table}
\end{table}

\subsubsection{\textbf{The influence of Modules}}
\label{sec:modules}
\renewcommand{\labelenumi}{\textbullet}
\begin{enumerate}[leftmargin=1em]
\item\textbf{w/o GE}: We remove the graph enhancement module. The model's average performance declines without graph enhancement, indicating that our graph enhancement module can effectively improve overall recommendation performance. We will further demonstrate the impact of graph enhancement on improving the recommendation performance for tail items in Section \ref{sec:long_tail}.
\item\textbf{w/o AL}: We remove the alignment for capturing commonalities module. On four datasets, the performance of the model without this module is significantly lower than that of \(\model\). This indicates that common information within modalities and between modalities and external behaviors is crucial for recommendation performance. Without the alignment process, the modality data of items might contain noise unrelated to the items themselves and a lot of noise unrelated to user behavior. Through alignment, we capture commonalities relevant to the recommendation scenario within modalities and between modalities and external behaviors, thereby achieving a denoising effect. This demonstrates the effectiveness of the alignment module.
\item\textbf{w/o UM}: We remove the user modality enhancement module. User modality enhancement is equally crucial for model performance, as the model's performance significantly declines without this component. Although it is possible to model the user's multimodal features by aggregating the multimodal features of the items the user has interacted with, this method has limitations. It only reflects the user's current multimodal preferences and lacks a comprehensive understanding of the user's latent interests and preferences. In contrast, leveraging contrastive learning between extended interest embeddings and explicit interaction embeddings helps to more comprehensively learn the user's multimodal preferences, thereby enhancing recommendation performance. We will further demonstrate the effectiveness of user modality augmentation through visualization in Section \ref{sec:visualization}.
\end{enumerate}

\begin{figure}
    \centering
    \begin{subfigure}[b]{0.22\textwidth}
        \includegraphics[width=\textwidth]{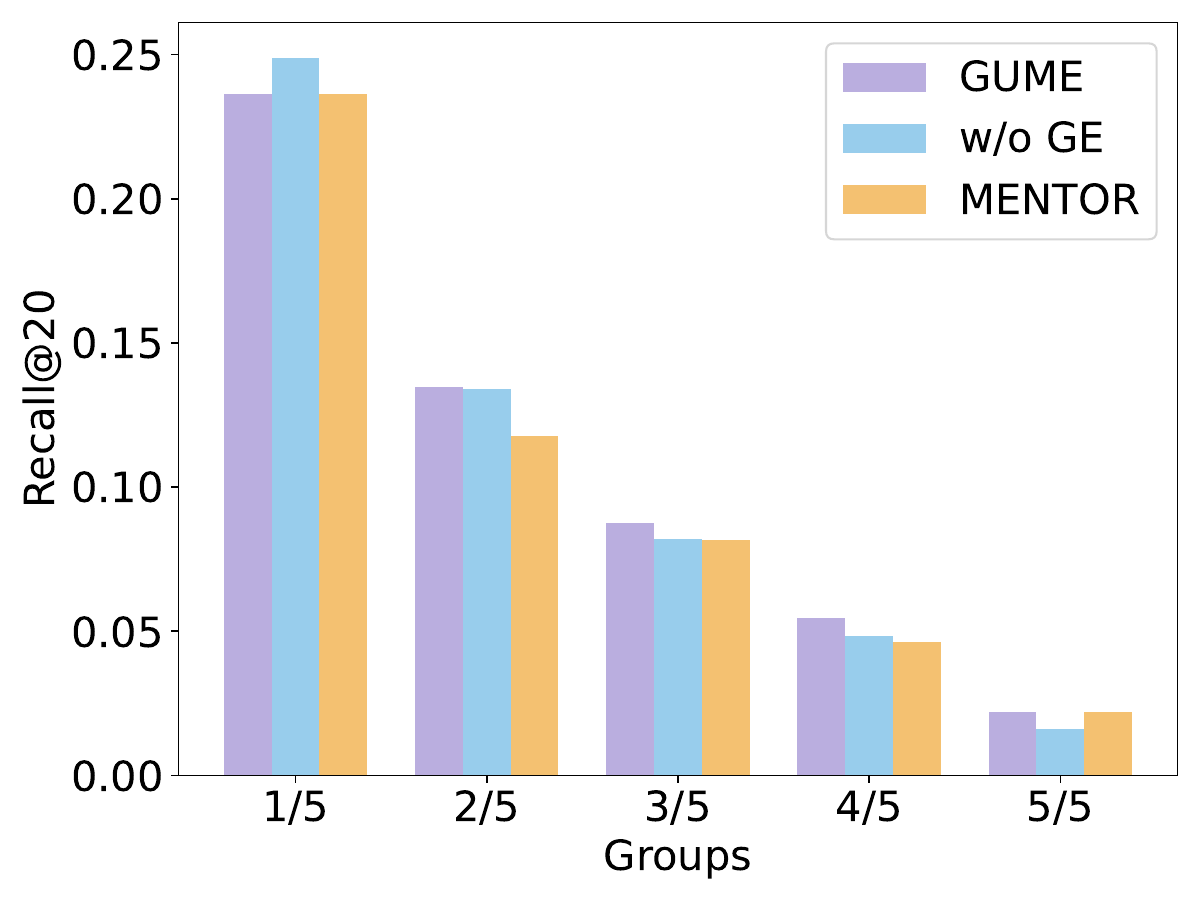}
        \caption{Recall@20}
        \label{fig:lt_clo_r}
    \end{subfigure}
    \hspace{0.01\textwidth}
    \begin{subfigure}[b]{0.22\textwidth}
        \includegraphics[width=\textwidth]{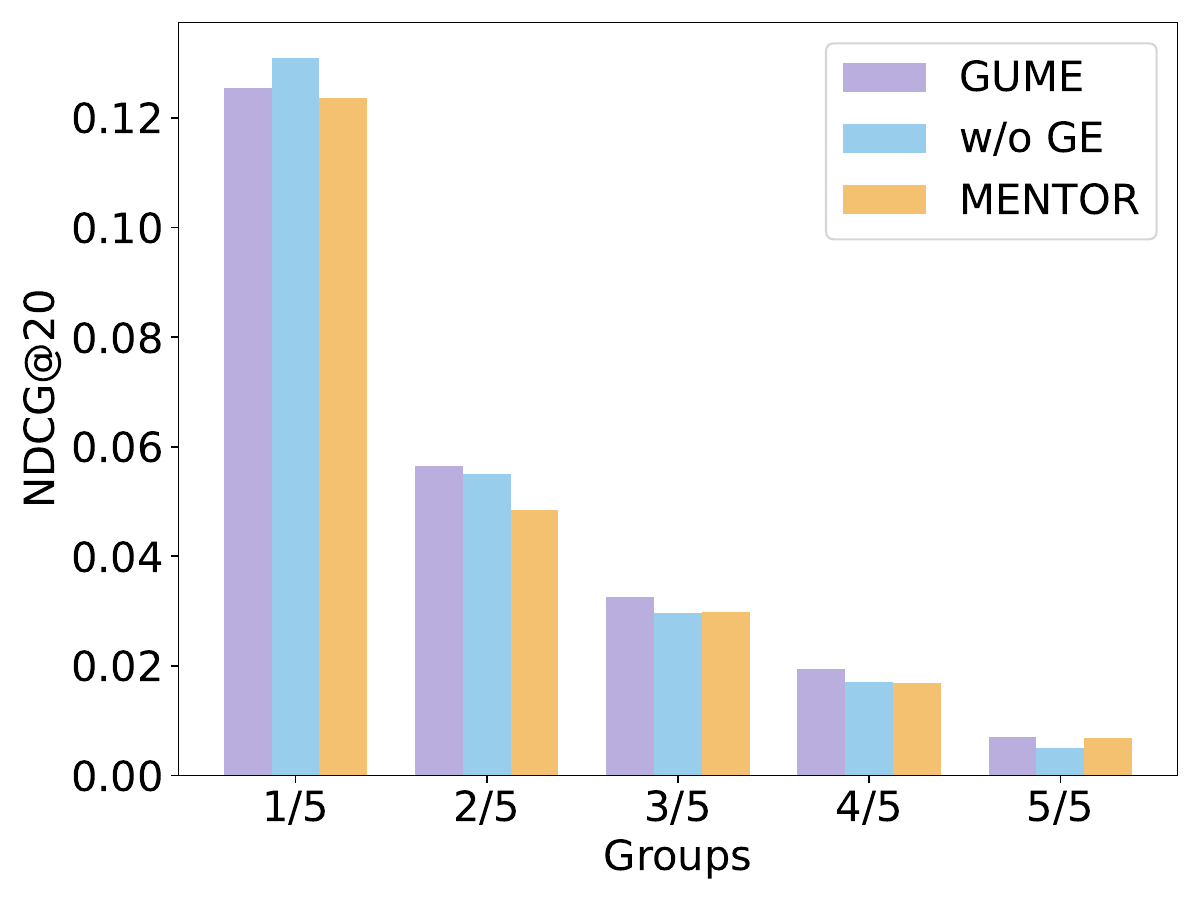}
        \caption{NDCG@20}
        \label{fig:lt_clo_n}
    \end{subfigure}
    \caption{Performance comparison of different item groups. We also compare the recommendation performance of long-tail items with MENTOR.}
    \label{fig:long_tail}
\end{figure}

\subsection{Comparisons on Tail Items Performance (RQ3)}
\label{sec:long_tail}
To validate whether enhancing the user-item graph based on multimodal similarity can improve the recommendation performance for tail items, we conducted experiments on the Clothing dataset. Specifically, we divided items into five equally sized groups according to the node degree in the user-item bipartite graph, as shown in Figure \ref{fig:long_tail}. In recommendation systems, 20\% of items account for 80\% of interactions. Therefore, we define the top 1/5 of items as head items, while the remaining 4/5 are defined as tail items. The larger the x-axis value, the lower the node degree, and the less popular the item. We compared the performance of \(\model\), w/o GE, and MENTOR. The results show that graph enhancement can improve the recommendation performance for tail items. Although removing graph enhancement can improve the recommendation performance for head items, the overall performance decreases due to the decline in tail item performance, which is consistent with the findings of GALORE \cite{luo2023improving}. Additionally, \(\model\) outperforms MENTOR for both head and tail items, indicating that our graph enhancement strategy effectively improves recommendation performance for long-tail distribution data.

\begin{figure}
    \centering
    \begin{subfigure}[b]{0.2\textwidth}
        \includegraphics[width=\textwidth]{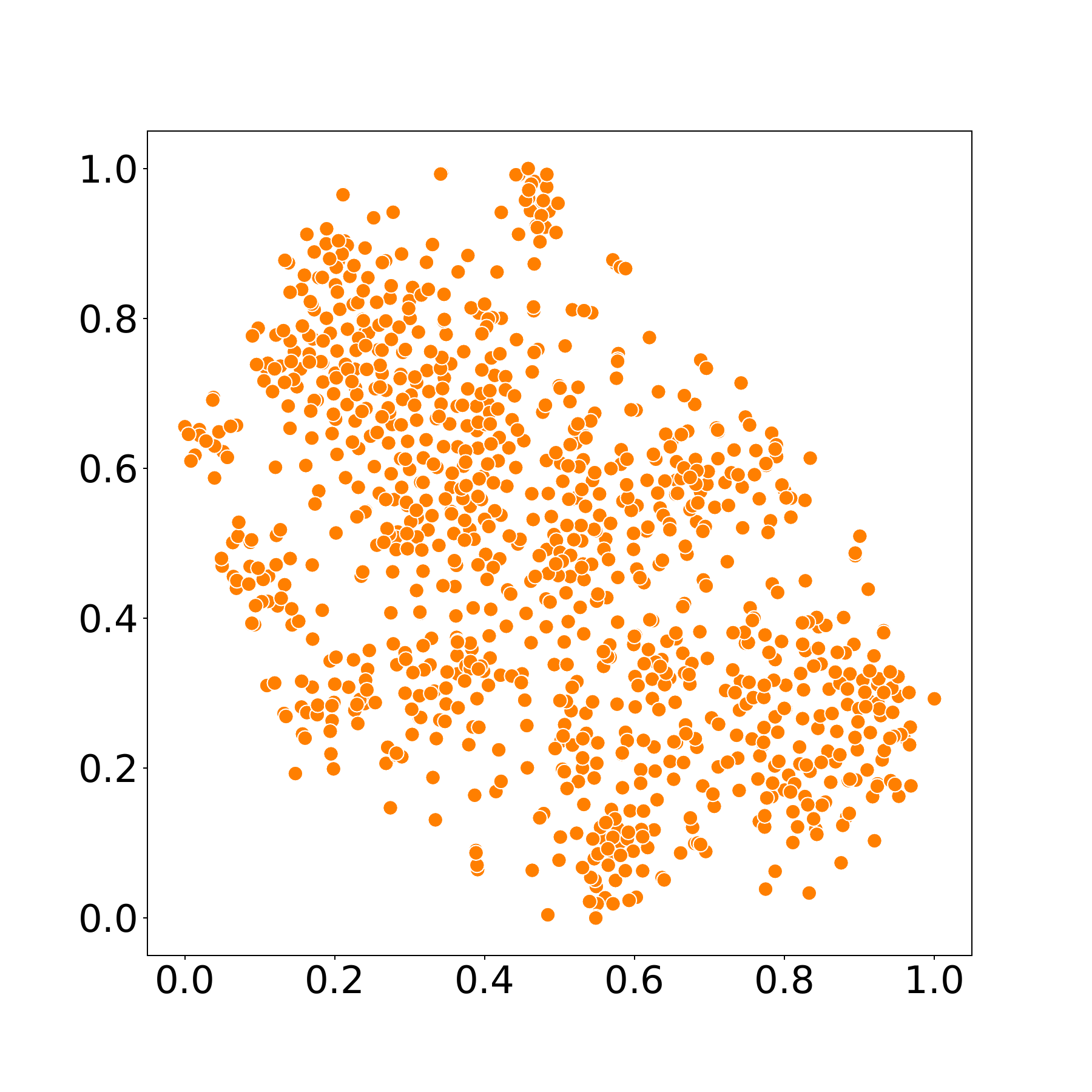}
        \caption{distribution on w/o AL}
        \label{fig:bad_side}
    \end{subfigure}
    ~
    \begin{subfigure}[b]{0.2\textwidth}
        \includegraphics[width=\textwidth]{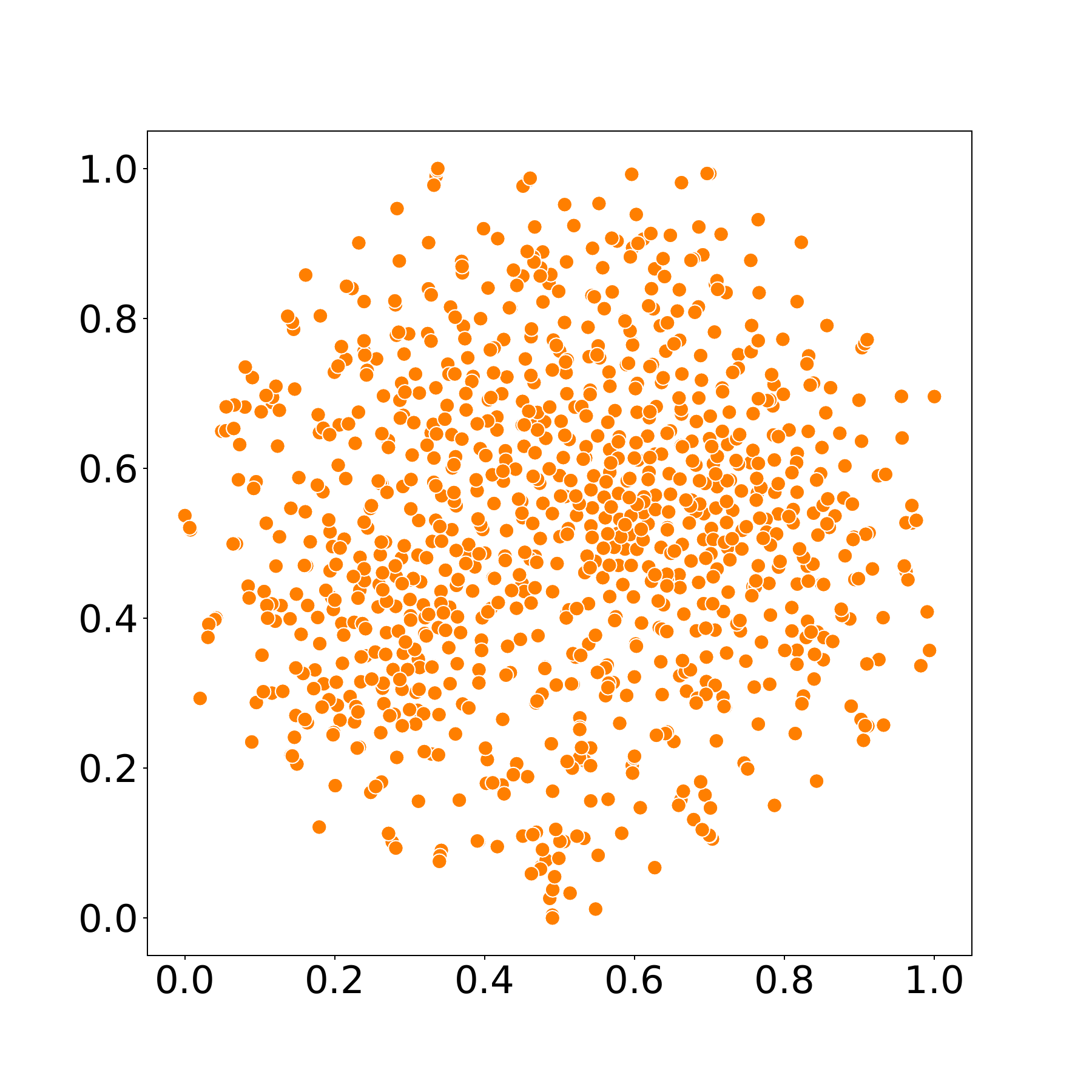}
        \caption{distribution on $\model$}
        \label{fig:good_side}
    \end{subfigure}
    \caption{The distribution of explicit interaction features for users, \(\bar{E}_{u,M}\). The left part of the figure shows the distribution without user modality enhancement, while the right part displays the distribution of \(\model\).}
    \label{fig:visualize_side}
\end{figure}

\subsection{Visualization Analysis (RQ4)}
\label{sec:visualization}
To further validate the effectiveness of the user modality enhancement component, we visualize the distribution of user modality representations within the Sports dataset. We compare two models, w/o UM and $\model$, as mentioned in section $\ref{sec:modules}$. Specifically, we randomly select 1000 user instances from the Sports dataset and employ t-SNE \cite{van2008visualizing} to map the user modality representations to two-dimensional space. The results, illustrated in figure \ref{fig:visualize_side}, show that the user modality distribution of $\model$ is more uniform, while the distribution of w/o UM is more dispersed. Previous research \cite{wang2022towards} has demonstrated that the uniformity of representation significantly influences recommendation performance. This explains why $\model$ is effective in enhancing user modality representation.

\begin{figure}
    \centering
    \begin{subfigure}[b]{0.2\textwidth}
        \includegraphics[width=\textwidth]{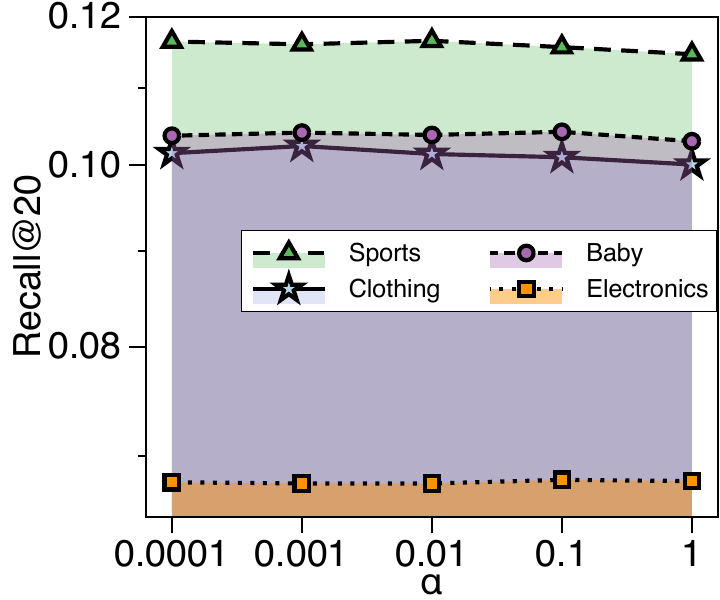}
        \caption{Recall@20}
        \label{fig:alpha_r20}
    \end{subfigure}
    \hspace{0.01\textwidth}
    \begin{subfigure}[b]{0.2\textwidth}
        \includegraphics[width=\textwidth]{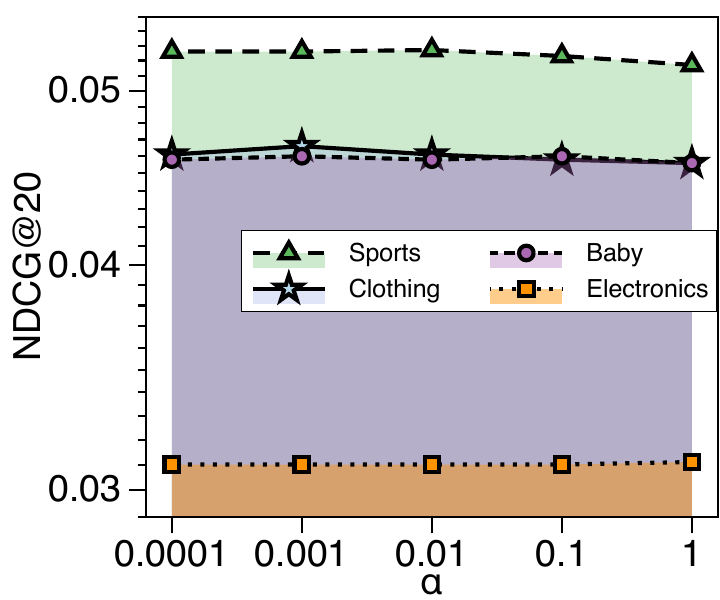}
        \caption{NDCG@20}
        \label{fig:alpha_n20}
    \end{subfigure}
    \caption{Effect of the balancing hyper-parameter $\alpha$}\label{fig:alpha}
\end{figure}
\subsection{Hyperparameter Sensitivity Study (RQ5)}
\subsubsection{\textbf{The balancing hyper-parameter $\alpha$}}
Figure \ref{fig:alpha} shows the variation of the balance hyperparameter $\alpha$ in the visual-textual alignment task within the range of \(\{1e^{-4}, 1e^{-3}, 1e^{-2}, 1e^{-1}, 1e^0\}\). Observations indicate that as $\alpha$ increases, performance initially improves and then declines. This suggests that jointly optimizing the visual-textual alignment task with the primary recommendation task can enhance performance, but if $\alpha$ is too large, the model may be misled by the visual-textual alignment task. Overall, there is no significant sharp rise or fall, indicating that our method is relatively insensitive to the choice of $\alpha$.

\begin{figure}[htbp]
    \centering
    \begin{subfigure}[b]{0.15\textwidth}
        \centering
        \includegraphics[width=\textwidth]{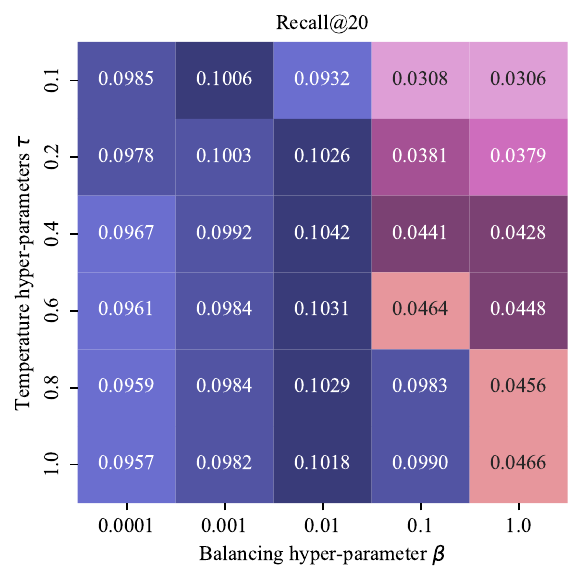}
        \caption{Baby}
        \label{fig:b_baby}
    \end{subfigure}
    \hspace{0.05\textwidth}
    \begin{subfigure}[b]{0.15\textwidth}
        \centering
        \includegraphics[width=\textwidth]{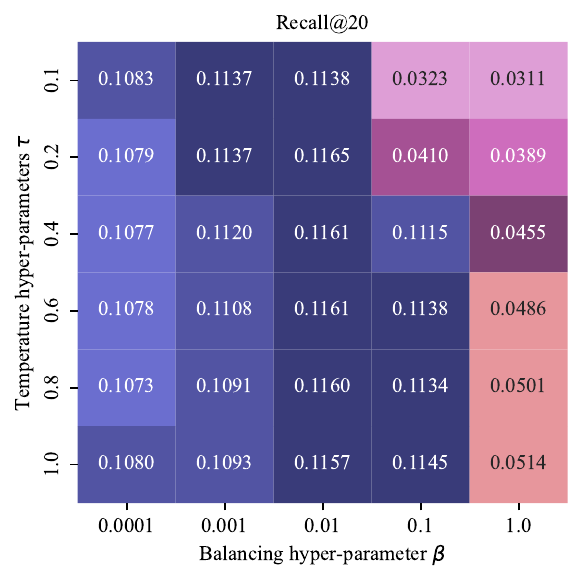}
        \caption{Sports}
        \label{fig:b_sports}
    \end{subfigure}
    
    \begin{subfigure}[b]{0.15\textwidth}
        \centering
        \includegraphics[width=\textwidth]{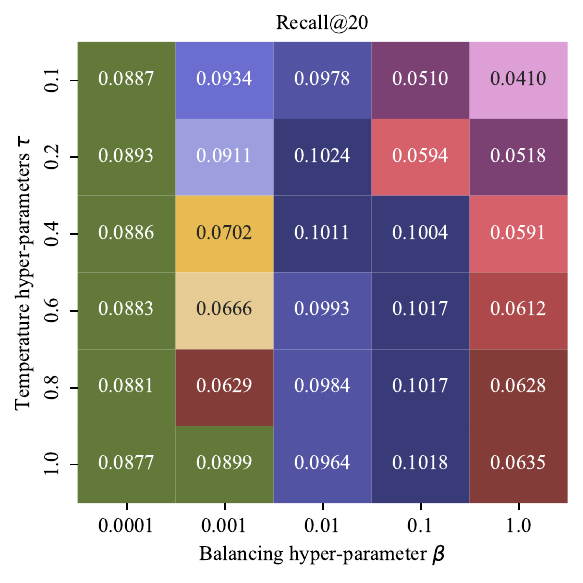}
        \caption{Clothing}
        \label{fig:b_clothing}
    \end{subfigure}
    \hspace{0.05\textwidth}
    \begin{subfigure}[b]{0.15\textwidth}
        \centering
        \includegraphics[width=\textwidth]{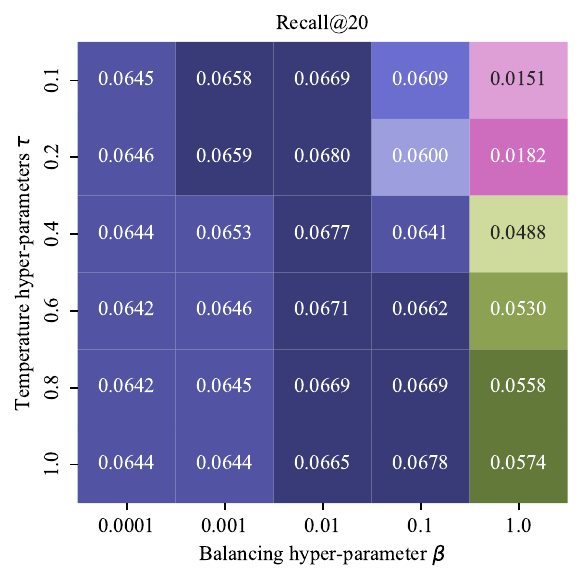}
        \caption{Electronics}
        \label{fig:b_electronics}
    \end{subfigure}
    
    \caption{The Recall@20 results for different pairs of \(\beta\) and \(\tau\).}
    \label{fig:beta and tau}
\end{figure}

\subsubsection{\textbf{The pair of hyper-parameters $\beta$ and $\tau$.}}
The behavior-modality alignment task is jointly controlled by the balance hyper-parameter $\beta$ and the temperature hyper-parameter $\tau$. We adjust $\beta$ within the range of \(\{1e^{-4}, 1e^{-3}, 1e^{-2}, 1e^{-1}, 1e^0\}\), and $\tau$ within the range of \{0.1, 0.2, 0.4, 0.6, 0.8, 1.0\}. As shown in Figure \ref{fig:beta and tau}, the model performs best on the sports, clothing, and electronics datasets when $\beta$ is set to 0.01 and $\tau$ is set to 0.2. On the Baby dataset, the best performance occurs when $\beta$ is 0.01 and $\tau$ is 0.4.

\begin{figure}[htbp]
    \centering
    \begin{subfigure}[b]{0.15\textwidth}
        \centering
        \includegraphics[width=\textwidth]{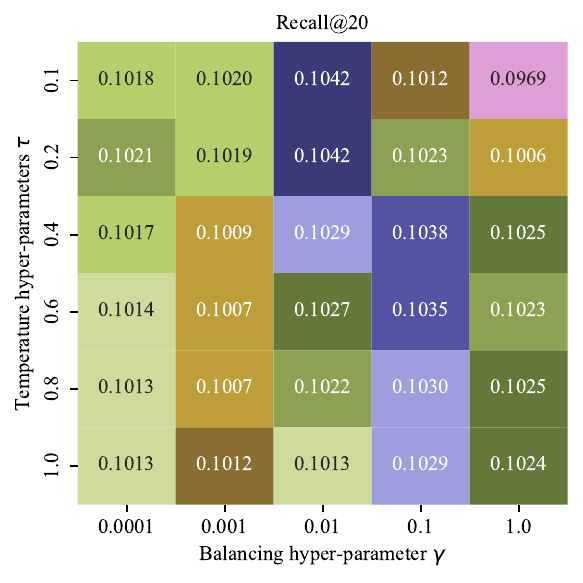}
        \caption{Baby}
        \label{fig:g_baby}
    \end{subfigure}
    \hspace{0.05\textwidth}
    \begin{subfigure}[b]{0.15\textwidth}
        \centering
        \includegraphics[width=\textwidth]{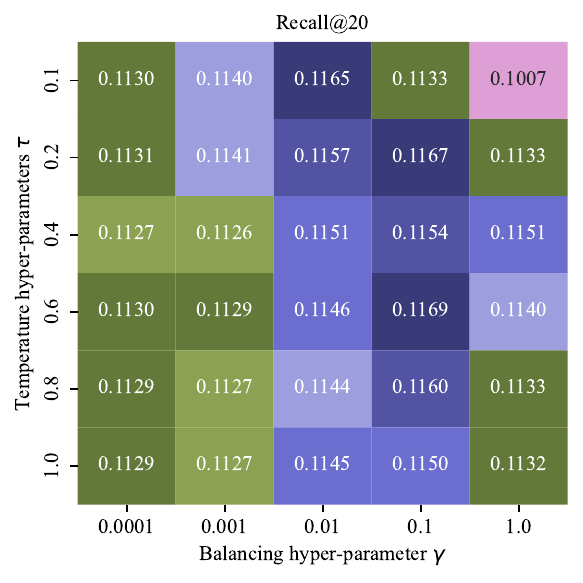}
        \caption{Sports}
        \label{fig:g_sports}
    \end{subfigure}
    
    \begin{subfigure}[b]{0.15\textwidth}
        \centering
        \includegraphics[width=\textwidth]{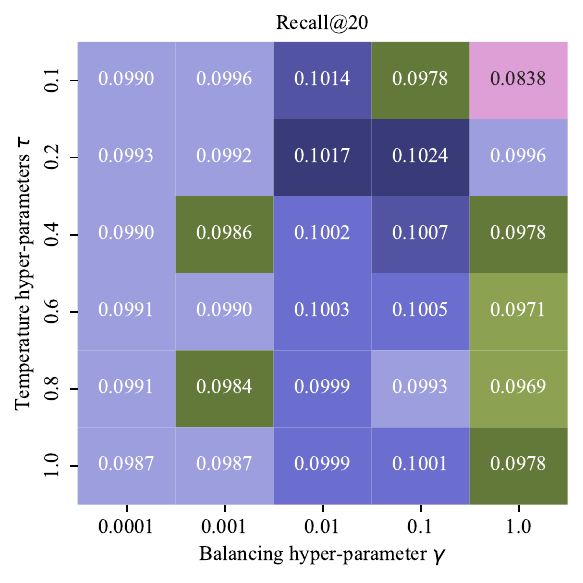}
        \caption{Clothing}
        \label{fig:g_clothing}
    \end{subfigure}
    \hspace{0.05\textwidth}
    \begin{subfigure}[b]{0.15\textwidth}
        \centering
        \includegraphics[width=\textwidth]{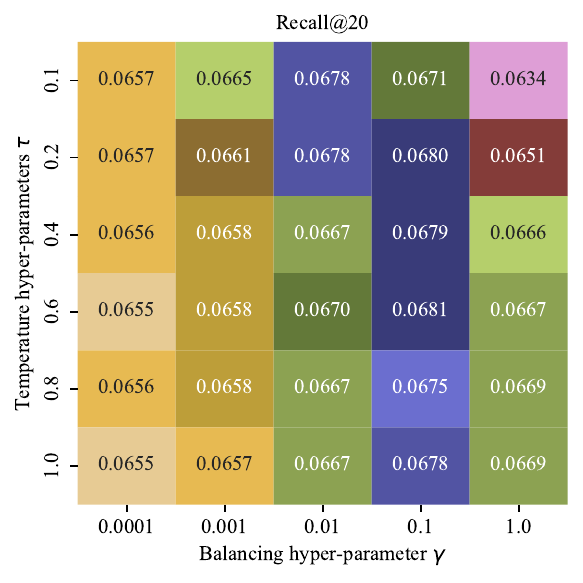}
        \caption{Electronics}
        \label{fig:g_electronics}
    \end{subfigure}
    
    \caption{The Recall@20 results for different pairs of \(\gamma\) and \(\tau\).}
    \label{fig:gamma and tau}
\end{figure}
\subsubsection{\textbf{The pair of hyper-parameters $\gamma$ and $\tau$.}}
The user modality enhancement task is jointly controlled by the balance hyper-parameter $\gamma$ and the temperature hyper-parameter $\tau$. We adjust $\gamma$ within the range of \(\{1e^{-4}, 1e^{-3}, 1e^{-2}, 1e^{-1}, 1e^0\}\) and $\tau$ within the range of \{0.1, 0.2, 0.4, 0.6, 0.8, 1.0\}. As shown in Figure \ref{fig:gamma and tau}, considering all metrics comprehensively, the model performs best on the smaller datasets (baby and sports) when $\gamma$ is set to 0.01 and $\tau$ is set to 0.1. On the larger datasets (clothing and electronics), the best performance occurs when $\gamma$ is set to 0.1 and $\tau$ is set to 0.2.

\section{Related Work}

\subsection{Multimedia Recommendation}
Recently, using multimodal information for recommendations has become a popular method to alleviate the data sparsity issue inherent in traditional recommendation systems. Typically, multimodal recommendations involve extracting features from multiple modalities using pre-trained neural networks, which are then integrated with behavioral features to better model user preferences. For example, VBPR \cite{he2016vbpr} enriches item representations by concatenating visual embeddings with ID embeddings. However, these item representations can be contaminated by features in the modal information that are irrelevant to user preferences. Inspired by graph convolutional networks, MMGCN \cite{wei2019mmgcn} leverages GCN to construct several modality-specific bipartite graphs of users and items, thereby uncovering hidden modality preferences within user-item interactions. GRCN \cite{wei2020grcn} reduces noise in the user-item bipartite graph by identifying and eliminating incorrect interaction data between users and items. However, these methods assign the same weight to each modality, overlooking the variations in user preferences for different modalities. To address this issue, DualGNN \cite{wang2021dualgnn} constructs a user co-occurrence graph and aggregates neighbor modality information based on this graph. MGCN \cite{yu2023MGCN} extracts modality preferences from user behavioral features and assigns modality weights based on these preferences. Additionally, utilizing item-item graphs can help achieve better item representations. LATTICE \cite{zhang2021LATTICE} and MICRO \cite{zhang2022micro} create item modality semantic graphs based on item modality information and aggregate them to capture latent item graphs. Self-supervised learning (SSL) has demonstrated remarkable performance in mitigating label dependency and addressing data sparsity issues. For instance, BM3 \cite{zhou2023bootstrap} does not introduce any auxiliary graphs but utilizes dropout techniques to generate contrastive views, thereby saving on model memory and computational costs. MENTOR \cite{xu2024mentor} employs aligned self-supervised tasks to synchronize multiple modalities while preserving interaction information.

\subsection{Long-tail Learning}
In the development of recommendation systems, the rapid increase in the number of items often exacerbates the long-tail effect, which in turn leads to cold start problems and a decline in recommendation quality \cite{yin2012challenging, volkovs2017dropoutnet}. To address this challenge, \cite{yin2012challenging} explored the application of random walk techniques on user-item graphs to find items that align with user interests, thereby addressing long-tail recommendations. DropoutNet \cite{volkovs2017dropoutnet} uses dropout techniques to train models, allowing it to make effective recommendation predictions based on content information when there is a lack of historical interaction data for users or items. Meta-learning focuses on using small amounts of training data to improve classification or regression performance, aligning well with the goals of recommendation systems and becoming a key solution for addressing long-tail recommendations. For example, MeLU \cite{lee2019melu} introduces a model-agnostic meta-learning algorithm (MAML) into recommendation systems to alleviate the cold start problem for users. Additionally, researchers have explored ways to enhance the recommendation performance of tail items. For example, TailNet \cite{liu2020long} introduces a preference mechanism that adjusts recommendation bias towards head or tail items by learning session representations and generating correction factors. MIRec \cite{zhang2021model} proposed a dual transfer learning strategy that facilitates knowledge transfer from head to tail items at both the model and item levels. GALORE \cite{luo2023improving} enhances tail item recommendations by using graph augmentation techniques, specifically by increasing the edge connections between head and tail items and selectively reducing the edge connections between users and head items.

\section{CONCLUSION}
In this paper, we propose a novel Graphs and User Modalities Enhancement framework, named $\model$, for long-tail multimodal recommendation. $\model$ enhances the user-item graph using multimodal similarities between items, improving the recommendation effectiveness for long-tail items. To effectively improve the generalization ability of user modality representations, $\model$ learns two independent representations: explicit interaction features and extended interest features. By maximizing the mutual information between these two features, the learned user modality features better adapt to changes in new products or user behavior. We also design two alignment tasks to denoise modality data from different perspectives. Experimental results on several widely-used datasets show that $\model$ significantly outperforms state-of-the-art multimodal recommendation methods.

\section{Ackowledgement}
This study is supported by grants from the Strategic Priority Research Program of the Chinese Academy of Sciences XDB38030300, the Postdoctoral Fellowship Program of CPSF (No.GZC20232736), the China Postdoctoral Science Foundation Funded Project \\(No.2023M743565), the Special Research Assistant Funded Project of the Chinese Academy of Sciences.

\printbibliography

@inproceedings{yu2023MGCN,
author = {Yu, Penghang and Tan, Zhiyi and Lu, Guanming and Bao, Bing-Kun},
title = {Multi-View Graph Convolutional Network for Multimedia Recommendation},
year = {2023},
isbn = {9798400701085},
publisher = {Association for Computing Machinery},
address = {New York, NY, USA},
url = {https://doi.org/10.1145/3581783.3613915},
doi = {10.1145/3581783.3613915},
booktitle = {Proceedings of the 31st ACM International Conference on Multimedia},
pages = {6576–6585},
numpages = {10},
location = {Ottawa ON,Canada},
series = {MM '23}
}

@article{yu2023ld4mrec,
  title={LD4MRec: Simplifying and Powering Diffusion Model for Multimedia Recommendation},
  author={Yu, Penghang and Tan, Zhiyi and Lu, Guanming and Bao, Bing-Kun},
  journal={arXiv preprint arXiv:2309.15363},
  year={2023}
}

@inproceedings{zhou2023bootstrap,
  title={Bootstrap latent representations for multi-modal recommendation},
  author={Zhou, Xin and Zhou, Hongyu and Liu, Yong and Zeng, Zhiwei and Miao, Chunyan and Wang, Pengwei and You, Yuan and Jiang, Feijun},
  booktitle={Proceedings of the ACM Web Conference 2023},
  pages={845--854},
  year={2023}
}

@inproceedings{chen2017attentive,
  title={Attentive collaborative filtering: Multimedia recommendation with item-and component-level attention},
  author={Chen, Jingyuan and Zhang, Hanwang and He, Xiangnan and Nie, Liqiang and Liu, Wei and Chua, Tat-Seng},
  booktitle={Proceedings of the 40th International ACM SIGIR conference on Research and Development in Information Retrieval},
  pages={335--344},
  year={2017}
}

@article{wei2017collaborative,
  title={Collaborative filtering and deep learning based recommendation system for cold start items},
  author={Wei, Jian and He, Jianhua and Chen, Kai and Zhou, Yi and Tang, Zuoyin},
  journal={Expert Systems with Applications},
  volume={69},
  pages={29--39},
  year={2017},
  publisher={Elsevier}
}

@article{deldjoo2016content,
  title={Content-based video recommendation system based on stylistic visual features},
  author={Deldjoo, Yashar and Elahi, Mehdi and Cremonesi, Paolo and Garzotto, Franca and Piazzolla, Pietro and Quadrana, Massimo},
  journal={Journal on Data Semantics},
  volume={5},
  pages={99--113},
  year={2016},
  publisher={Springer}
}

@inproceedings{yu2022SimGCL,
  title={Are graph augmentations necessary? simple graph contrastive learning for recommendation},
  author={Yu, Junliang and Yin, Hongzhi and Xia, Xin and Chen, Tong and Cui, Lizhen and Nguyen, Quoc Viet Hung},
  booktitle={Proceedings of the 45th international ACM SIGIR conference on research and development in information retrieval},
  pages={1294--1303},
  year={2022}
}

@inproceedings{sarwar2001item,
  title={Item-based collaborative filtering recommendation algorithms},
  author={Sarwar, Badrul and Karypis, George and Konstan, Joseph and Riedl, John},
  booktitle={Proceedings of the 10th international conference on World Wide Web},
  pages={285--295},
  year={2001}
}

@incollection{rendle2012bpr,
  title={BPR: Bayesian personalized ranking from implicit feedback},
  author={Rendle, Steffen and Freudenthaler, Christoph and Gantner, Zeno and Schmidt-Thieme, Lars},
  journal={arXiv preprint arXiv:1205.2618},
  year={2012}
}

@inproceedings{he2016vbpr,
  title={VBPR: visual bayesian personalized ranking from implicit feedback},
  author={He, Ruining and McAuley, Julian},
  booktitle={Proceedings of the AAAI conference on artificial intelligence},
  volume={30},
  number={1},
  year={2016}
}

@inproceedings{wei2019mmgcn,
  title={MMGCN: Multi-modal graph convolution network for personalized recommendation of micro-video},
  author={Wei, Yinwei and Wang, Xiang and Nie, Liqiang and He, Xiangnan and Hong, Richang and Chua, Tat-Seng},
  booktitle={Proceedings of the 27th ACM international conference on multimedia},
  pages={1437--1445},
  year={2019}
}

@inproceedings{he2020lightgcn,
  title={Lightgcn: Simplifying and powering graph convolution network for recommendation},
  author={He, Xiangnan and Deng, Kuan and Wang, Xiang and Li, Yan and Zhang, Yongdong and Wang, Meng},
  booktitle={Proceedings of the 43rd International ACM SIGIR conference on research and development in Information Retrieval},
  pages={639--648},
  year={2020}
}

@article{zhang2022micro,
  title={Latent structure mining with contrastive modality fusion for multimedia recommendation},
  author={Zhang, Jinghao and Zhu, Yanqiao and Liu, Qiang and Zhang, Mengqi and Wu, Shu and Wang, Liang},
  journal={IEEE Transactions on Knowledge and Data Engineering},
  year={2022},
  publisher={IEEE}
}

@inproceedings{zhou2023mmrec,
  title={Mmrec: Simplifying multimodal recommendation},
  author={Zhou, Xin},
  booktitle={Proceedings of the 5th ACM International Conference on Multimedia in Asia Workshops},
  pages={1--2},
  year={2023}
}

@article{tao2022SLMRec,
  title={Self-supervised learning for multimedia recommendation},
  author={Tao, Zhulin and Liu, Xiaohao and Xia, Yewei and Wang, Xiang and Yang, Lifang and Huang, Xianglin and Chua, Tat-Seng},
  journal={IEEE Transactions on Multimedia},
  year={2022},
  publisher={IEEE}
}

@article{xu2024mentor,
  title={MENTOR: Multi-level Self-supervised Learning for Multimodal Recommendation},
  author={Xu, Jinfeng and Chen, Zheyu and Yang, Shuo and Li, Jinze and Wang, Hewei and Ngai, Edith C-H},
  journal={arXiv preprint arXiv:2402.19407},
  year={2024}
}

@inproceedings{luo2023improving,
  title={Improving long-tail item recommendation with graph augmentation},
  author={Luo, Sichun and Ma, Chen and Xiao, Yuanzhang and Song, Linqi},
  booktitle={Proceedings of the 32nd ACM International Conference on Information and Knowledge Management},
  pages={1707--1716},
  year={2023}
}

@inproceedings{liu2021interest,
  title={Interest-aware message-passing GCN for recommendation},
  author={Liu, Fan and Cheng, Zhiyong and Zhu, Lei and Gao, Zan and Nie, Liqiang},
  booktitle={Proceedings of the web conference 2021},
  pages={1296--1305},
  year={2021}
}

@article{baltruvsaitis2018multimodal,
  title={Multimodal machine learning: A survey and taxonomy},
  author={Baltru{\v{s}}aitis, Tadas and Ahuja, Chaitanya and Morency, Louis-Philippe},
  journal={IEEE transactions on pattern analysis and machine intelligence},
  volume={41},
  number={2},
  pages={423--443},
  year={2018},
  publisher={IEEE}
}

@inproceedings{glorot2010understanding,
  title={Understanding the difficulty of training deep feedforward neural networks},
  author={Glorot, Xavier and Bengio, Yoshua},
  booktitle={Proceedings of the thirteenth international conference on artificial intelligence and statistics},
  pages={249--256},
  year={2010},
  organization={JMLR Workshop and Conference Proceedings}
}

@article{kingma2014adam,
  title={Adam: A method for stochastic optimization},
  author={Kingma, Diederik P and Ba, Jimmy},
  journal={arXiv preprint arXiv:1412.6980},
  year={2014}
}

@inproceedings{wei2020grcn,
  title={Graph-refined convolutional network for multimedia recommendation with implicit feedback},
  author={Wei, Yinwei and Wang, Xiang and Nie, Liqiang and He, Xiangnan and Chua, Tat-Seng},
  booktitle={Proceedings of the 28th ACM international conference on multimedia},
  pages={3541--3549},
  year={2020}
}

@inproceedings{zhang2021LATTICE,
  title={Mining latent structures for multimedia recommendation},
  author={Zhang, Jinghao and Zhu, Yanqiao and Liu, Qiang and Wu, Shu and Wang, Shuhui and Wang, Liang},
  booktitle={Proceedings of the 29th ACM international conference on multimedia},
  pages={3872--3880},
  year={2021}
}

@article{yin2012challenging,
  title={Challenging the long tail recommendation},
  author={Yin, Hongzhi and Cui, Bin and Li, Jing and Yao, Junjie and Chen, Chen},
  journal={arXiv preprint arXiv:1205.6700},
  year={2012}
}

@article{volkovs2017dropoutnet,
  title={Dropoutnet: Addressing cold start in recommender systems},
  author={Volkovs, Maksims and Yu, Guangwei and Poutanen, Tomi},
  journal={Advances in neural information processing systems},
  volume={30},
  year={2017}
}

@inproceedings{lee2019melu,
  title={Melu: Meta-learned user preference estimator for cold-start recommendation},
  author={Lee, Hoyeop and Im, Jinbae and Jang, Seongwon and Cho, Hyunsouk and Chung, Sehee},
  booktitle={Proceedings of the 25th ACM SIGKDD International Conference on Knowledge Discovery \& Data Mining},
  pages={1073--1082},
  year={2019}
}

@inproceedings{liu2020long,
  title={Long-tail session-based recommendation},
  author={Liu, Siyi and Zheng, Yujia},
  booktitle={Proceedings of the 14th ACM Conference on Recommender Systems},
  pages={509--514},
  year={2020}
}

@inproceedings{zhang2021model,
  title={A model of two tales: Dual transfer learning framework for improved long-tail item recommendation},
  author={Zhang, Yin and Cheng, Derek Zhiyuan and Yao, Tiansheng and Yi, Xinyang and Hong, Lichan and Chi, Ed H},
  booktitle={Proceedings of the web conference 2021},
  pages={2220--2231},
  year={2021}
}

@inproceedings{zhou2023tale,
  title={A tale of two graphs: Freezing and denoising graph structures for multimodal recommendation},
  author={Zhou, Xin and Shen, Zhiqi},
  booktitle={Proceedings of the 31st ACM International Conference on Multimedia},
  pages={935--943},
  year={2023}
}

@article{van2008visualizing,
  title={Visualizing data using t-SNE.},
  author={Van der Maaten, Laurens and Hinton, Geoffrey},
  journal={Journal of machine learning research},
  volume={9},
  number={11},
  year={2008}
}

@inproceedings{wang2022towards,
  title={Towards representation alignment and uniformity in collaborative filtering},
  author={Wang, Chenyang and Yu, Yuanqing and Ma, Weizhi and Zhang, Min and Chen, Chong and Liu, Yiqun and Ma, Shaoping},
  booktitle={Proceedings of the 28th ACM SIGKDD conference on knowledge discovery and data mining},
  pages={1816--1825},
  year={2022}
}

@article{chen2009knn,
  title={Fast Approximate kNN Graph Construction for High Dimensional Data via Recursive Lanczos Bisection.},
  author={Chen, Jie and Fang, Haw-ren and Saad, Yousef},
  journal={Journal of Machine Learning Research},
  volume={10},
  number={9},
  year={2009}
}

@article{wang2021dualgnn,
  title={Dualgnn: Dual graph neural network for multimedia recommendation},
  author={Wang, Qifan and Wei, Yinwei and Yin, Jianhua and Wu, Jianlong and Song, Xuemeng and Nie, Liqiang},
  journal={IEEE Transactions on Multimedia},
  volume={25},
  pages={1074--1084},
  year={2021},
  publisher={IEEE}
}

@inproceedings{ft1,
  title={Self-optimizing feature generation via categorical hashing representation and hierarchical reinforcement crossing},
  author={Ying, Wangyang and Wang, Dongjie and Liu, Kunpeng and Sun, Leilei and Fu, Yanjie},
  booktitle={2023 IEEE International Conference on Data Mining (ICDM)},
  pages={748--757},
  year={2023},
  organization={IEEE}
}

@article{ft2,
  title={Unsupervised Generative Feature Transformation via Graph Contrastive Pre-training and Multi-objective Fine-tuning},
  author={Ying, Wangyang and Wang, Dongjie and Hu, Xuanming and Zhou, Yuanchun and Aggarwal, Charu C and Fu, Yanjie},
  journal={arXiv preprint arXiv:2405.16879},
  year={2024}
}

@inproceedings{ft5,
  title={Traceable automatic feature transformation via cascading actor-critic agents},
  author={Xiao, Meng and Wang, Dongjie and Wu, Min and Qiao, Ziyue and Wang, Pengfei and Liu, Kunpeng and Zhou, Yuanchun and Fu, Yanjie},
  booktitle={Proceedings of the 2023 SIAM International Conference on Data Mining (SDM)},
  pages={775--783},
  year={2023},
  organization={SIAM}
}

@article{ft4,
  title={Traceable group-wise self-optimizing feature transformation learning: A dual optimization perspective},
  author={Xiao, Meng and Wang, Dongjie and Wu, Min and Liu, Kunpeng and Xiong, Hui and Zhou, Yuanchun and Fu, Yanjie},
  journal={ACM Transactions on Knowledge Discovery from Data},
  volume={18},
  number={4},
  pages={1--22},
  year={2024},
  publisher={ACM New York, NY}
}

@article{ft3,
  title={Enhancing Tabular Data Optimization with a Flexible Graph-based Reinforced Exploration Strategy},
  author={Huang, Xiaohan and Wang, Dongjie and Ning, Zhiyuan and Qiao, Ziyue and Long, Qingqing and Zhu, Haowei and Wu, Min and Zhou, Yuanchun and Xiao, Meng},
  journal={arXiv preprint arXiv:2406.07404},
  year={2024}
}

@article{fs1,
  title={Feature Selection as Deep Sequential Generative Learning},
  author={Ying, Wangyang and Wang, Dongjie and Chen, Haifeng and Fu, Yanjie},
  journal={arXiv preprint arXiv:2403.03838},
  year={2024}
}

@article{fs2,
  title={Neuro-Symbolic Embedding for Short and Effective Feature Selection via Autoregressive Generation},
  author={Gong, Nanxu and Ying, Wangyang and Wang, Dongjie and Fu, Yanjie},
  journal={arXiv preprint arXiv:2404.17157},
  year={2024}
}

@inproceedings{fs3,
  title={Beyond discrete selection: Continuous embedding space optimization for generative feature selection},
  author={Xiao, Meng and Wang, Dongjie and Wu, Min and Wang, Pengfei and Zhou, Yuanchun and Fu, Yanjie},
  booktitle={2023 IEEE International Conference on Data Mining (ICDM)},
  pages={688--697},
  year={2023},
  organization={IEEE}
}

@article{fsa1,
  title={Enhanced Gene Selection in Single-Cell Genomics: Pre-Filtering Synergy and Reinforced Optimization},
  author={Zhang, Weiliang and Meng, Zhen and Wang, Dongjie and Wu, Min and Liu, Kunpeng and Zhou, Yuanchun and Xiao, Meng},
  journal={arXiv preprint arXiv:2406.07418},
  year={2024}
}

@article{fsa2,
  title={FedGCS: A Generative Framework for Efficient Client Selection in Federated Learning via Gradient-based Optimization},
  author={Ning, Zhiyuan and Tian, Chunlin and Xiao, Meng and Fan, Wei and Wang, Pengyang and Li, Li and Wang, Pengfei and Zhou, Yuanchun},
  journal={arXiv preprint arXiv:2405.06312},
  year={2024}
}

@inproceedings{qq3,
  title={Graph structural-topic neural network},
  author={Long, Qingqing and Jin, Yilun and Song, Guojie and Li, Yi and Lin, Wei},
  booktitle={Proceedings of the 26th ACM SIGKDD International Conference on Knowledge Discovery \& Data Mining},
  pages={1065--1073},
  year={2020}
}

@inproceedings{qq2,
  title={Unveiling Delay Effects in Traffic Forecasting: A Perspective from Spatial-Temporal Delay Differential Equations},
  author={Long, Qingqing and Fang, Zheng and Fang, Chen and Chen, Chong and Wang, Pengfei and Zhou, Yuanchun},
  booktitle={Proceedings of the ACM on Web Conference 2024},
  pages={1035--1044},
  year={2024}
}

@inproceedings{qq1,
  title={HGK-GNN: Heterogeneous Graph Kernel based Graph Neural Networks},
  author={Long, Qingqing and Xu, Lingjun and Fang, Zheng and Song, Guojie},
  booktitle={Proceedings of the 27th ACM SIGKDD Conference on Knowledge Discovery \& Data Mining},
  pages={1129--1138},
  year={2021}
}

@inproceedings{m3,
  title={Needed: Introducing hierarchical transformer to eye diseases diagnosis},
  author={Ye, Xu and Xiao, Meng and Ning, Zhiyuan and Dai, Weiwei and Cui, Wenjuan and Du, Yi and Zhou, Yuanchun},
  booktitle={Proceedings of the 2023 SIAM International Conference on Data Mining (SDM)},
  pages={667--675},
  year={2023},
  organization={SIAM}
}

@inproceedings{m2,
  title={Expert knowledge-guided length-variant hierarchical label generation for proposal classification},
  author={Xiao, Meng and Qiao, Ziyue and Fu, Yanjie and Du, Yi and Wang, Pengyang and Zhou, Yuanchun},
  booktitle={2021 IEEE International Conference on Data Mining (ICDM)},
  pages={757--766},
  year={2021},
  organization={IEEE}
}

@article{m1,
  title={Hierarchical interdisciplinary topic detection model for research proposal classification},
  author={Xiao, Meng and Qiao, Ziyue and Fu, Yanjie and Dong, Hao and Du, Yi and Wang, Pengyang and Xiong, Hui and Zhou, Yuanchun},
  journal={IEEE Transactions on Knowledge and Data Engineering},
  volume={35},
  number={9},
  pages={9685--9699},
  year={2023},
  publisher={IEEE}
}
% \bibliography{ref}
% \bibliographystyle{ACM-Reference-Format}

\clearpage
\appendix

\end{document}